\documentclass[aps,prd,superscriptaddress,showpacs,preprintnumbers,nofootinbib]{revtex4}
\usepackage{graphicx,xcolor}
\usepackage{wrapfig}

\usepackage{amsmath}
\usepackage{amsfonts}
\usepackage{amssymb}
\def\Journal#1#2#3#4{{#1} {\bf #2}, #3 (#4)}

\def\PRD{{\em Phys. Rev.} D}
\def\ZPC{{\em Z. Phys.} C}
\def\EPG{{\em Eur.Phys.J.} C}

\def\be{\begin{equation}}
\def\ee{\end{equation}}
\def\bea{\begin{eqnarray}}
\def\eea{\end{eqnarray}}

\begin{document}

\title{Three Pomerons vs D\O~and TOTEM Data}  

\preprint{JLAB-THY-12-1667}

\author{V. A. Petrov}
  \email{Vladimir.Petrov@ihep.ru}
  \affiliation{Division of Theoretical Physics, Institute for High Energy Physics,\\
142281 Protvino,  RUSSIA }
 \author{A. Prokudin}
  \email{prokudin@jlab.org}  
  \affiliation{Jefferson Lab, 12000 Jefferson Avenue, Newport News, VA 23606}
   
\date{\today}

\begin{abstract} 
%%%%%%%%%%%%%%%%%%%%%%%%%%%%%%%%%%%%%%%%%%%%%%%%%%%%%%%%%%%%
This note highlights the predictive power of the 3-component Pomeron model \cite{threepomerons,Petrov:2002nt},
designed by the authors ten years ago with a partial account of the multiplicity of Pomerons in QCD.
The model is put to the test by comparing its predictions with the recent data from D\O~ \cite{Abazov:2012qb} and TOTEM \cite{Antchev:2011zz,Antchev:2011vs,Antchev:2012} Collaborations at 1.96 and 7 TeV, respectively.
We also compare model predictions for inelastic cross section to experimental measurement by TOTEM, CMS, ALICE, and ATLAS
collaborations.
It is shown that the D\O~\cite{Abazov:2012qb} data are perfectly predicted by the model. Total, elastic, and inelastic cross section predictions are in agreement with the measurements by TOTEM~\cite{Antchev:2011zz}, CMS~\cite{CMS1,CMS}, ALICE~\cite{Abelev:2012sja} and ATLAS~\cite{Aad:2011eu} Collaborations. Differential cross section data at 7 TeV show slight disagreement with predictions
of the model in high-$t$ region. Discussions on the origin of the disagreement and conclusions are presented. 
\end{abstract} 

\maketitle

\section{INTRODUCTION}

Publication of the D\O~ data \cite{Abazov:2012qb}
on elastic proton-antiproton scattering at $\sqrt{s}=1.96$ TeV and the TOTEM data \cite{Antchev:2011zz,Antchev:2011vs,Antchev:2012} on elastic proton-proton differential scattering cross-section
at $\sqrt{s}=7$ TeV and inelastic cross section by TOTEM collaboration Ref.~\cite{Antchev:2011zz}, CMS Collaboration Ref.~\cite{CMS1,CMS},  ALICE Collaboration Ref.~\cite{Abelev:2012sja}, and ATLAS Collaboration Ref.~\cite{Aad:2011eu} has triggered a great deal of interest in the diffractive community \cite{Bourrely:2012hp,Soffer:2012pq,Grau:2012wy,Selyugin:2012gv,Nemes:2012cp,Merino:2012qc,Godizov:2012fy,Ryskin:2012ry,Schegelsky:2011aa,Lengyel:2012sw,Troshin:2012nu,Gotsman:2012rq,
Block:2012yx,Block:2011vz,Block:2012nj,Donnachie:2011aa,Kopeliovich:2012yy}. 

One of the striking features of the new data is that the predictions of the models 
~\cite{Petrov:2002nt,Block:2011uy,Bourrely:2002wr,Islam:2009zza,Jenkovszky:2011bt,Kaspar:2011zz} considered in Ref.~\cite{Antchev:2011zz},
while often close to the measured experimental values fail to predict the TOTEM data in detail. 
A compilation of these models was made in Ref.~\cite{Kaspar:2011zz}.
 
In this article we would like to consider the model of Ref.~\cite{threepomerons} where predictions were made for LHC at $\sqrt{s}=14$ TeV
and to check the predictions of this model for D\O~ at $\sqrt{s}=1.96$ TeV and LHC (TOTEM) at $\sqrt{s}=7$ TeV.
We are going to use the parameters of the model fixed in Ref.~\cite{threepomerons} to ensure all our results
are genuine predictions.
One of the intrinsic ingredients of the model of Ref.~\cite{threepomerons} is a phenomenological account of a well established fact
that in theories with asymptotic freedom there exist multiple leading Regge trajectories with intercepts that form infinite set of numbers
(independent of the coupling constant) accumulating near some minimum limit, see
 Ref.~\cite{Lovelace}.

In massless QCD the intercepts of the leading Pomeron set of trajectories behave like (see the second item in Ref.~\cite{Lovelace} )
 $\alpha_{ {\mathbb P}_k}$ - 1= $\Delta_{ {\mathbb P}_k}\propto\frac{const}{k}$ at $k \rightarrow\infty $.
Up to now no more detailed results derived rigorously from  QCD, at least for the leading Pomeron trajectories. 
We should mention a recent attempt ~\cite{Kancheli:2011xz} based on some extra assumptions.
Explicit formulas were obtained recently in  Ref.~\cite{Godizov:2009du} for (massless) quark-antiquark trajectories.

For the time being we have decided to mimic the proliferation of Pomerons with a few (two, three, four, ...) tentative trajectories
with all parameters to be phenomenologically defined by fitting the data.
The best fit chooses the 3-Pomeron option. It is worth mentioning a later attempt ~\cite{Ellis:2008yp}
to use a many-component
QCD Pomeron for description of the DIS data from HERA. Phenomenologically a two-Pomeron 
model with one ``soft'' and one ``hard'' Pomerons  was suggested earlier in 
Ref.~\cite{Donnachie:1998gm}.

\section{THE MODEL}
Let us first briefly describe the model of the elastic scattering amplitude from Refs.~\cite{threepomerons,Petrov:2002nt}. The nuclear amplitude as function of the 
impact parameter $b$ 
is written in the eikonal form as
\begin{equation}
T^N(s,\vec b)=\frac{e^{2i\delta (s,\vec b)}-1}{2i}\; ,
\label{eq:ampl}
\end{equation}
where the eikonal $\delta(s,b)$ is approximated by single-reggeon exchanges
\be
\delta_{pp}^{\bar p p}(s,b) = \delta^+_{{\mathbb P}}(s,b)  
\mp \delta^-_{\mathbb
O}(s,b)+\delta^+_{
f}(s,b)\mp \delta^-_{\omega}(s,b).
\label{eq:modeleik}
\ee

We refer the reader to the original literature for details; let us simply 
recall that here $\delta^+_{{\mathbb P}}(s,b)$ is the Pomeron contributions and superscript `$+$' denotes C-even trajectories (the Pomeron trajectories have 
quantum numbers $0^+J^{++}$), while `$-$' denotes  C-odd trajectories. 
$\delta^-_{\mathbb O}(s,b)$ is the Odderon contribution (i.e. the C-odd partner of
the Pomeron whose quantum numbers are $0^-J^{--}$); $\delta^+_{ f}$ and 
$\delta^-_{\omega}(s,b)$ are the contributions of secondary Reggeons, $f$
as representative of the $C=+1$ families and $\omega$ of the $C=-1$.

This approximation for the eikonal is similar to 
the impact parameter approximation in quantum mechanics with single-reggeon exchanges in the role of potentials.
Actually this assumption is generic for many models in the field. From the standpoint of the complex $J$-plane, the eikonal has to have not only Regge poles but, generally, more complicated singularities.
Nonetheless, we believe that the pole approximation for the eikonal reasonably reflects the
gross features of the diffractive hadron scattering. In our model we use linear trajectories and this could be justified if
one considers only low-$t$ region.  Our linear Pomeron trajectories
can go below the line $J=1$ at sufficiently large $-t$, however, Regge trajectories in QCD are non-linear at negative $t$, see Refs.~\cite{Petrov}.
Pomeron trajectories most probably always lie higher than $J=1$, see Ref.~\cite{Kirschner}. If there is a disagreement with data, we have
to pay closer attention to such details.

As was said, the  effective number of Pomerons was found to be three, and
the pomeron contribution $\delta^+_{{\mathbb P}}(s,b)$ can be, accordingly, represented as the sum of three contributions:
\begin{equation}
\delta^+_{{\mathbb P}}(s,b) = \delta^+_{{\mathbb P}_1}(s,b)+
\delta^+_{{\mathbb P}_2}(s,b)+
\delta^+_{{\mathbb P}_3}(s,b) \; ,
\label{eq:pomerons}
\end{equation}
each of those having a particular Regge trajectory. Other models include complicated form-factors \cite{Donnachie:2011aa} or a different type of singularity for the Pomeron \cite{Bourrely:2012hp}. 

\begin{wrapfigure}{r}{0.55\textwidth}
\begin{center}
%\begin{figure}
\vspace{-30pt}
	\includegraphics[width=0.45\textwidth,bb= 10 140 540 660]{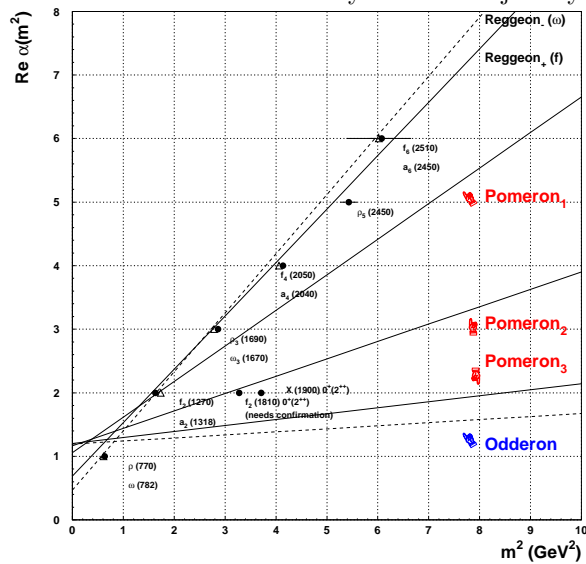}
	\caption{Regge trajectories from Ref.~\cite{threepomerons}.\label{fig:1}}
%\end{figure}
\vspace{50pt}
\end{center}
\end{wrapfigure}

Let us recall that if the form of Eq.~\eqref{eq:pomerons} is indeed manifested in nature, then we expect to have particles lying on the corresponding trajectories. The model predicts $M_{glueball}^2 = 1.68, 3.05, 8.51$ (GeV$^2$) for $0^+2^{++}$ state.
The trajectories from Ref.~\cite{threepomerons} are presented in Fig.~\ref{fig:1}. In fact one of the $0^+2^{++}$ candidates is situated very close to trajectory of 
${\mathbb P}_2$.

In Ref.~\cite{threepomerons} this three-component Pomeron model
was successfully used for description of
high energy $pp$ and $\bar pp$  data in the region 
of large momentum transfers
$0.01 \le |t| \le 14.5$ GeV$^2$. In Ref.~\cite{Petrov:2002nt} the model was extended
to the region of small momentum transfers $0 \le |t| \le 0.01$ GeV$^2$. In order to do this we have to account 
for  the Coulomb interaction. The standard way is
to represent the whole scattering amplitude $T(s,t)$ which is dominated by the Coulomb force at
low momentum transfers and by the hadronic force at higher momentum transfers
as

\be
T(s,t)=T^N(s,t)+e^{i\alpha\Phi}T^C(s,t),
\ee
where we normalize the scattering amplitude so that 
\vskip 0.5cm
\begin{align}
\displaystyle \frac{d\sigma}{dt} = \frac{\vert T(s,t) \vert ^2}{16\pi s^2},
\end{align}
and the Born Coulomb amplitude for $pp$ and $\bar pp$  scattering is
\be
\displaystyle T^C(s,t) = \mp\frac{8\pi \alpha s}{|t|}.
\ee
The upper (lower) sign corresponds
to the scattering of particles with the same (opposite) charges. $T^N(s,t)$ stands for purely strong interaction amplitude, and
the phase $\Phi$ depends generally on energy,  momentum transfer and on the properties 
of $T^N$.  We considered three different forms of the phase $\Phi$ calculated earlier 
by West and Yennie~\cite{West} ,  Cahn~\cite{Cahn} and 
Selyugin~\cite{Selyugin} and showed in Ref.~\cite{Petrov:2002nt} that all three phases lead 
to a good description of the
low-$t$ data.

In order to relate $t$- and $b$-spaces one proceeds via Fourier-Bessel
transforms
 
\begin{align}
\displaystyle \hat f(t)= 4 \pi s\int_{0}^{\infty} db^2\ J_0(b\sqrt{-t}) f(b)\; , \\
\displaystyle f(b)= \frac{1}{16 \pi s}\int_{-\infty}^{0} dt\ J_0(b\sqrt{-t}) \hat f(t) \; .
\label{eq:fb}
\end{align}

Crossing symmetry is restored by replacing
$s \rightarrow s e^{-i\pi/2}$. We introduce the dimensionless variable
\be
\tilde s = \frac{s}{s_0}e^{-i\frac{\pi}{2}}\ ,\; s_{0} = 1 \; {\rm(GeV^2)}\; ,
\ee
in terms of which we give each $C+$ and $C-$ contribution
with an appropriate signature factor in the form

\begin{align}
\displaystyle \delta^{\pm} (s,b)=\mathcal{C} \frac{c}{s_0}
\tilde s^{\alpha(0)-1}\frac{e^{-\frac{b^2}{\rho^2}}}{4\pi \rho^2}\ ,
 \; {\rm with}\;
\rho^2 = 4\alpha'(0) \ln\tilde s+r^2\ ,  \label{eq:eikonalform}
\end{align}
where $\mathcal{C} = i$ for $C+$ and $\mathcal{C} = 1$ for $C-$.

For the cross-sections we use the following normalizations:Total and elastic cross-sections are normalized such that

\begin{align}
\sigma_{tot} = \frac{1}{s} \Im {\rm m} T(s,t=0), \;\;
\sigma_{elastic} = 4 \pi \int_{0}^{\infty}db^2 \vert T(s,b) \vert^2. 
\end{align}

 %%%%%%%%%%%%%%%%%%%%%%%%%%%%%%%%%%%%%%%%%%%%%%%%
\begin{figure}[h]
\centering
  \begin{tabular}{c@{\hspace*{5mm}}c}
    \includegraphics[width=0.45\textwidth,bb= 20 160 520 640]{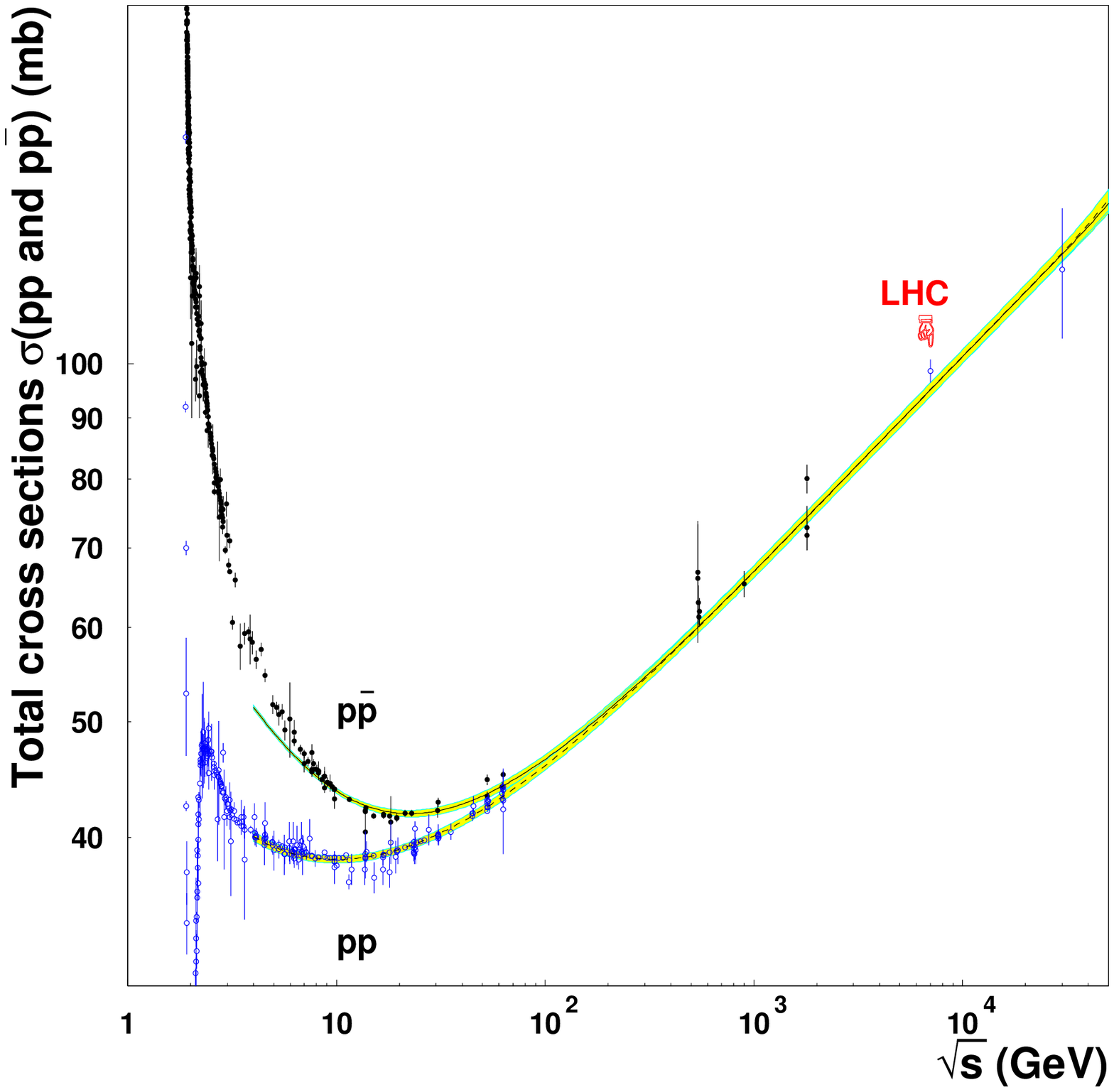}
    &
    \includegraphics[width=0.45\textwidth,bb= 20 160 520 640]{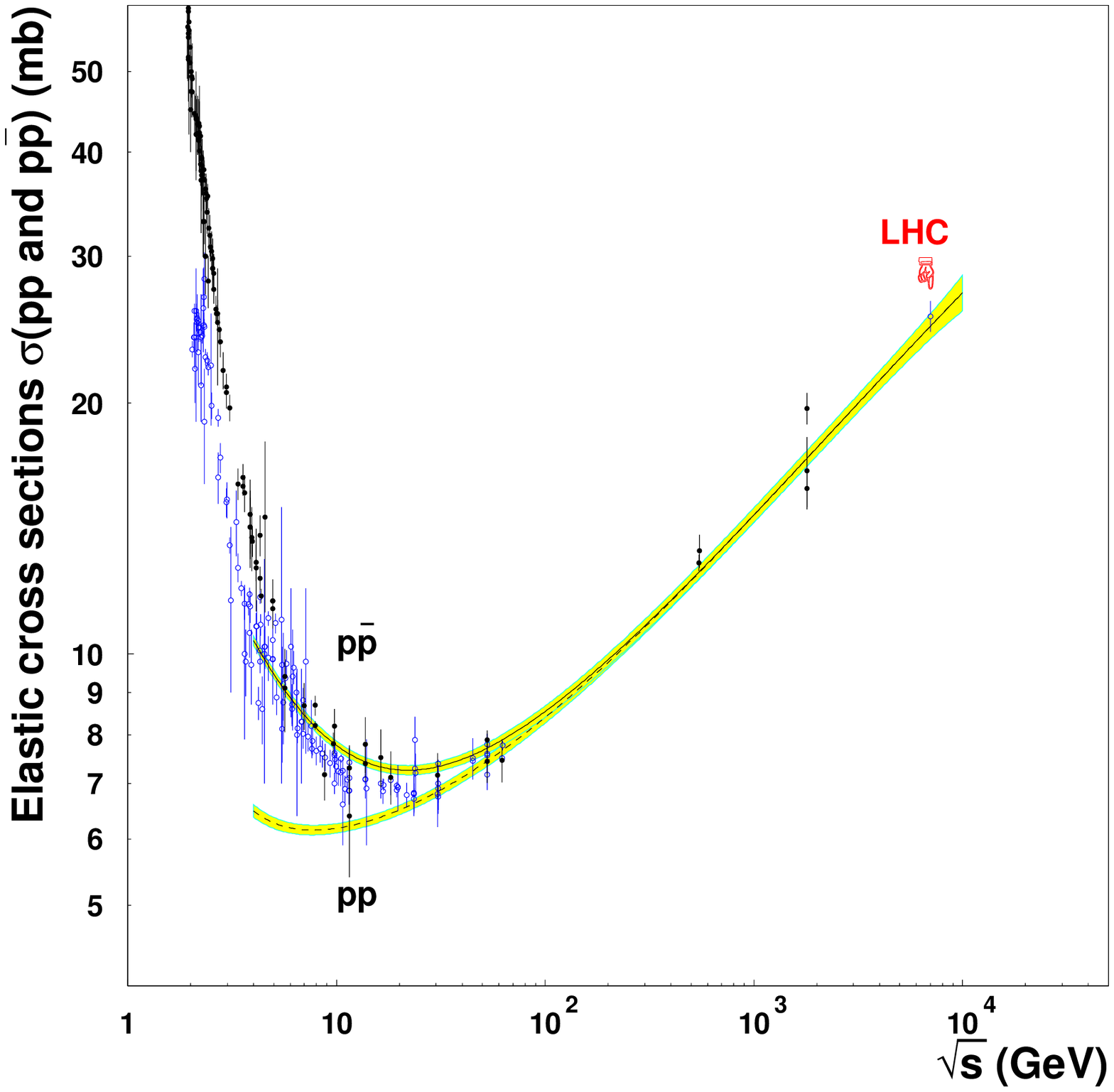}  
  \\
  (a) & (b)
  \end{tabular}
	\caption{Total (a) and elastic (b) cross sections for $pp$ and $p\bar p$ scattering. The yellow band corresponds to the error propagation of parameter uncertainties of the model. \label{fig:totelastic}}
\end{figure}
 %%%%%%%%%%%%%%%%%%%%%%%%%%%%%%%%%%%%%%%%%%%%%%%%
 
\section{RESULTS AND DISCUSSIONS}
 
In Ref.~\cite{threepomerons}, the adjustable parameters were fitted 
over a set of 982 $pp$ and $\bar p p$ data points \cite{Data} of both forward observables 
(total cross-sections $\sigma_{tot}$, and $\rho$, ratios of the real to 
the imaginary part of the amplitude) in the range $8\le\sqrt{s}\le 1800$ GeV 
and angular distributions
($\frac{d\sigma}{dt}$) in the ranges $23\le\sqrt{s}\le 1800$ GeV,
$0.01\le |t|\le 14$ GeV$^2$. A good $\chi^2/d.o.f. = 2.6$ (we did not consider systematic errors in normalizations of data sets
and thus assumed that the value of $\chi^2/d.o.f.$ to be satisfactory) was obtained and
the parameters are given in Appendix ~\ref{appendix} Table~\ref{tab:1}. A set of the data
 including the Coulomb region 
which consists of 2158 points \cite{Data} was considered in Ref.~\cite{Petrov:2002nt} and it was shown that
Coulomb-nuclear interference   well   described the data. We also found that the Coulomb phases from Ref.~\cite{Cahn} and  Ref.~\cite{Selyugin} gave a
good account of the data in the lowest-$t$ region.

\begin{figure}[ht]
	\includegraphics[width=0.55\textwidth,bb= 20 160 520 640]{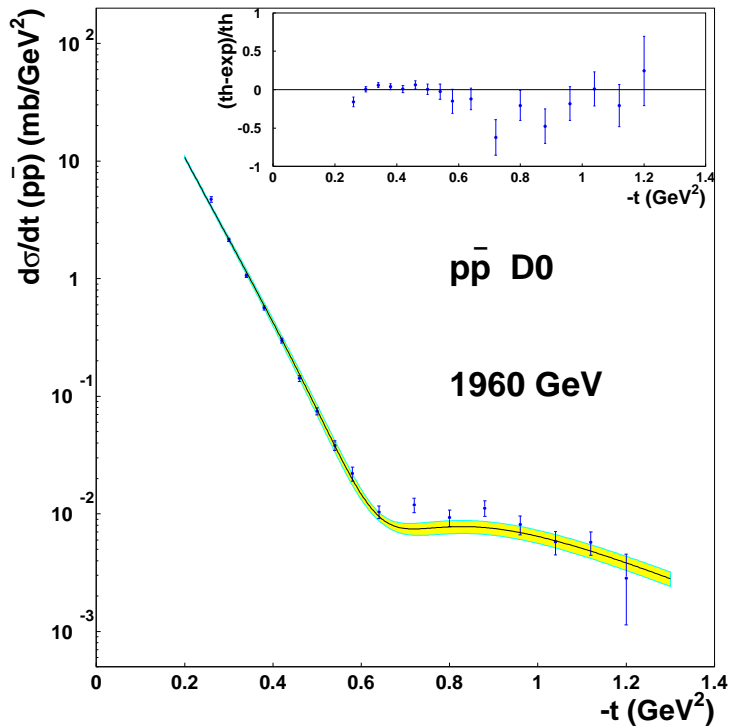}
	\caption{Differential cross section for $p\bar p$ scattering at 1.96 TeV and comparison with the
D\O~ data. The yellow band corresponds to the error propagation of parameter uncertainties of the model. The upper panel shows the variance of the model predictions and the data. \label{fig:difpbarpd0}}
\end{figure}

The total and elastic cross sections are presented in Fig. \ref{fig:totelastic}. Note that in contrast to Ref.~\cite{threepomerons} we included
the error corridor as explained in Appendix~\ref{appendix}. In Fig.~\ref{fig:totelastic} we also plot the experimental values of the total and elastic cross sections found by the TOTEM. As one can see, the model predictions are very close to the 
data. We show $p\bar p$  angular distributions over the full range of $|t|$  and in comparison with the D\O~ data at 1.96 TeV in Fig.~\ref{fig:difpbarpd0}. One can see that the data are predicted perfectly well by the model.

 %%%%%%%%%%%%%%%%%%%%%%%%%%%%%%%%%%%%%%%%%%%%%%%%
 %%%%%%%%%%%%%%%%%%%%%%%%%%%%%%%%%%%%%%%%%%%%%%%%

\begin{figure}[h]
\centering
  \begin{tabular}{c@{\hspace*{5mm}}c}
    \includegraphics[width=0.48\textwidth,bb= 20 160 520 640]{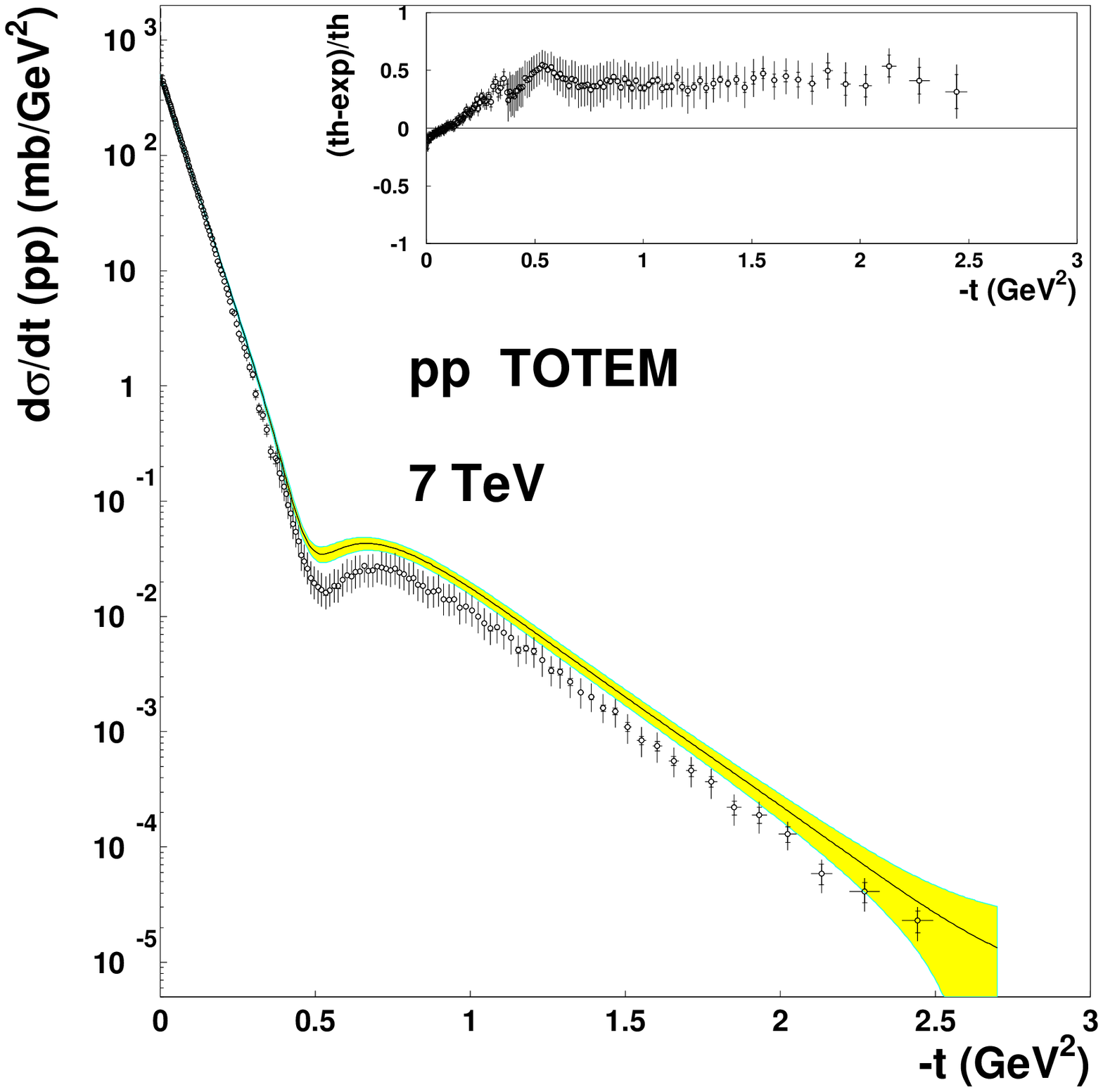}
    &
    \includegraphics[width=0.48\textwidth,bb= 20 160 520 640]{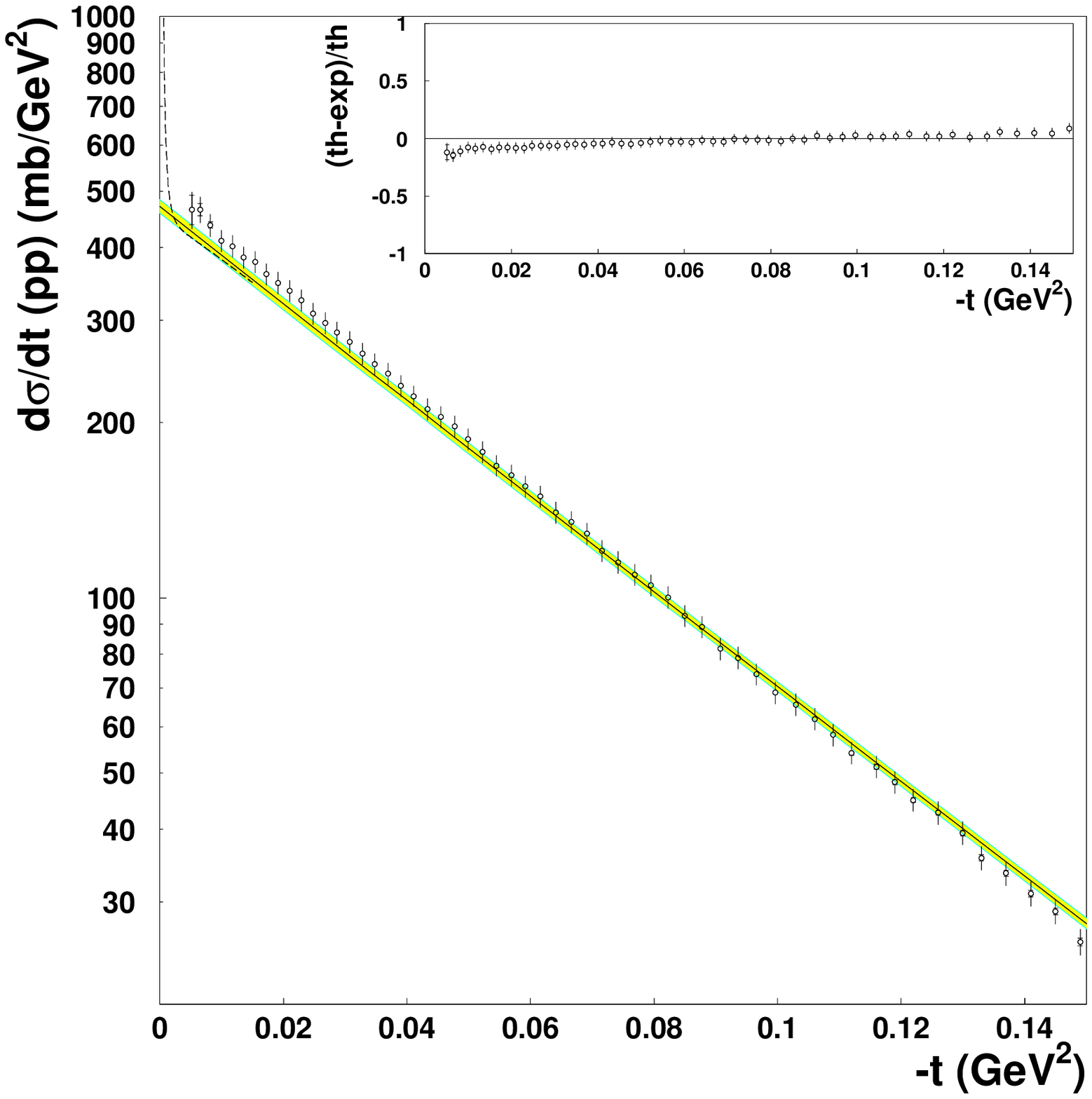}  
  \\
  (a) & (b)
  \end{tabular}
	\caption{Differential cross section for $pp$ scattering at 7 TeV in comparison with the
TOTEM data over whole (a) and small-$t$ (b) regions. The yellow band corresponds to the error propagation of parameter uncertainties of the model. Upper panel shows the variance of the model predictions and the data. Cross section with Ñoulomb-nuclear interference is shown by the dashed line in (b). 
 \label{fig:diffpplhc}}
\end{figure}

We show $pp$ angular distributions over the full range of $|t|$ in comparison with the TOTEM data at 7 TeV in Fig.~\ref{fig:diffpplhc} (a). Considering that the measurement spans
$8$ orders of magnitude, the match of the prediction and the measurements is fairly satisfactory. The model has some deviation from the
data in the dip region, around $|t|\ge |t_{dip}| = 0.52$ GeV$^2$, however this region is the most subtle one, so we leave its description to a future analysis.
In the small-$t$ region (see Fig.~\ref{fig:diffpplhc} (b)) the comparison of our predictions and the measured cross section remains   fair, and the data and predictions differ by less than 10\%. However we must observe that the slope of the measured cross section is higher than our prediction (see  Fig.~\ref{fig:diffpplhc} (b)). Comparison of predicted values and measured ones by TOTEM are presented in Table.~\ref{tab:2}. One can see that the model 
predicts the general trend of the data quite well. There are, however, nonnegligible deviations, in particular,
the total cross section is slightly underestimated, the slope of the elastic differential cross section and 
 $d\sigma_{el}/dt|_{t=0}$ are slightly smaller than measured values. 
Inelastic cross section at 7 TeV was measured by the TOTEM Collaboration Ref.\cite{Antchev:2011zz,Antchev:2012}, the CMS Collaboration Ref.~\cite{CMS1,CMS}, the  ALICE Collaboration Ref.\cite{Abelev:2012sja}, and the ATLAS Collaboration Ref.~\cite{Aad:2011eu} while
our model prediction is $\sigma_{inelastic} = 70.34 \pm 2.11$ mb and 
$\sigma_{inelastic}/\sigma_{tot} = 0.74$ (see Table~\ref{tab:2}). 
This ratio is  close to its experimental value $\sigma_{inelastic}/\sigma_{tot} \approx 0.72$. 

Given the quality of the comparison we are confident that,on the whole, our model has proved its predictive power. 
Detailed analysis of the impact of the TOTEM data on the parameters of the model will be presented elsewhere.

 %%%%%%%%%%%%%%%%%%%%%%%%%%%%%%%%%%%%%%%%%%%%%%%%
 %%%%%%%%%%%%%%%%%%%%%%%%%%%%%%%%%%%%%%%%%%%%%%%%

\begin{table}[h]
\begin{center} 
\begin{tabular}{llc} 
\hline
\multicolumn{2}{c}{Experimental result}  & Model prediction  \\
\hline
$\sigma_{tot}$ [mb], TOTEM, \cite{Antchev:2012}& $98.58\pm2.23$ &$95.06\pm1.26$ \\ 
$\sigma_{elastic}$ [mb], TOTEM, \cite{Antchev:2012}& $25.42\pm1.07$ &$24.72\pm0.85$ \\ 
$\sigma_{inelastic}$ [mb], TOTEM, \cite{Antchev:2011vs}& $73.5\pm0.6^{+1.8}_{-1.3}$ &$70.34 \pm 2.11$ \\ 
$\sigma_{inelastic}$ [mb], CMS, \cite{CMS}& $64.5\pm3.0\pm1.5$ &$70.34 \pm 2.11$ \\ 
$\sigma_{inelastic}$ [mb], CMS, \cite{CMS1}& $68\pm2\pm2.4\pm4$ &$70.34 \pm 2.11$ \\ 
$\sigma_{inelastic}$ [mb], ATLAS, \cite{Aad:2011eu}& $69.4 \pm 2.4\pm 6.9$ &$70.34 \pm 2.11$ \\ 
$\sigma_{inelastic}$ [mb], ALICE, \cite{Abelev:2012sja}& $73.2^{+2.0}_{-4.6}\pm2.6$  &$70.34 \pm 2.11$ \\ 
$d\sigma_{el}/dt|_{t=0}$ [mb/GeV$^2$], TOTEM,  \cite{Antchev:2012}& $506.4\pm23.0$ &$470.9\pm12.5$ \\ 
$B$ [mb/GeV$^2$] [GeV$^{-2}$], TOTEM, \cite{Antchev:2012} &$19.89\pm0.27$ &$19.32$ \\ 
$B_{(|t|=0.4\; GeV^2)}$   TOTEM, [GeV$^{-2}$], \cite{Antchev:2011zz}&$23.6\pm0.5\pm05$ &$22.1$ \\ 
$|t_{dip}|$   [GeV$^{2}$], TOTEM, \cite{Antchev:2011zz}&$0.53\pm0.01\pm0.01$ &$0.52$ \\ 
\hline
\end{tabular}
\end{center}
\caption{Comparison of experimental measurement by TOTEM collaboration Ref.\cite{Antchev:2011zz,Antchev:2012}, CMS Collaboration Ref.~\cite{CMS1,CMS},  ALICE Collaboration Ref.\cite{Abelev:2012sja}, and ATLAS Collaboration Ref.~\cite{Aad:2011eu} and model predictions Ref.~\cite{threepomerons}
at $\sqrt{s}$= 7 TeV. 
\label{tab:2}}
\end{table} 
 
  %%%%%%%%%%%%%%%%%%%%%%%%%%%%%%%%%%%%%%%%%%%%%%%%
\section*{CONCLUSIONS}

The above analysis shows that the D\O~ data \cite{Abazov:2012qb} on the total and differential cross-sections support our predictions.
Total, elastic, and inelastic cross section predictions are in agreement with the measurements by TOTEM~\cite{Antchev:2011zz,Antchev:2012}, CMS~\cite{CMS1,CMS}, ALICE~\cite{Abelev:2012sja} and ATLAS~\cite{Aad:2011eu}.
The TOTEM \cite{Antchev:2012} results on differential elastic cross-section, while being in agreement with the predictions of such general characteristics as the forward slope and the position of the dip reveal a moderate but non-negligible discrepancy with the observed $t$-dependence at large $t$. Nonetheless, the disagreement with the TOTEM \cite{Antchev:2012} is, to our mind, far from being deadly.
We believe that the model should be improved as, for example, in view of possible future measurements of the diffractive central production
of Higgs and other states, there is a need for a good model of diffraction, whose parameters constitute an important element of the corresponding modeling.
We have also to add that the very model contains, in its theoretical framework, 
a lot of potential sources for further refinement.

\section{Acknowledgements}
 We are grateful to Roman Ryutin and Anton Godizov for useful discussions 
 of the subject of this note and to Michael Pennington for critical reading 
of the manuscript. Authored by a Jefferson Science Associate, LLC under U.S. DOE Contract 
No. DE-AC05-06OR23177. The U.S. Government retains a non-exclusive, 
paid-up, irrevocable, 
world-wide license to publish
or reproduce this manuscript for U.S. Government purposes.
%\newpage

\appendix
\section{\label{appendix}}
In order to estimate the error corridor of our predictions from Ref.~\cite{threepomerons} we use the 
following method.Generically for a measured function $f(x,\{a\})$ that depends on variable $x$ and a set of parameters $\{a\} = \{a_1,...,a_N\}$, error propagation reads:
\begin{align}
 (\Delta f(x,\{a\}))^2 = {\sum_{i=1}^N  \sum_{j=1}^N C_{ij} \frac{\partial f}{\partial a_i} \sigma_i  \frac{\partial f}{\partial a_j}}\sigma_j   \; ,
\label{eq:error}
\end{align}
where $\sigma_i ,\sigma_j$ are one sigma errors on parameters (in our case we use those from from Table~\ref{tab:1}) and $C_{ij}$ is the correlation coefficient for parameters $i$ and $j$ . For totally uncorrelated parameters we have $C_{ij} = \delta_{ij}$ and the formula reduces to the standard one:
\begin{align}
 \left(\Delta f(x,\{a\})\right)^2 = \sum_{i=1}^N   \left(\frac{\partial f}{\partial a_i}\right)^2 \sigma_i^2     \; .
\end{align}

We extract the error correlation matrix from  Ref.~\cite{threepomerons} and apply Eq.~\eqref{eq:error} to all measured quantities plotted in this note.

{\small
\begin{table}[h]
\begin{center}
\begin{tabular}{lcllcl}
\hline
 \multicolumn{3}{c}{\bf Pomeron$_{\bf 1}$} &  \multicolumn{3}{c}{\bf Odderon}   \\
\hline
$\Delta_{{\mathbb P}_1}$&$=$ & $0.0578\pm0.0020$ &  $\Delta_{{\mathbb O}}$&$=$ & $0.19200\pm0.0025$ \\
$c_{{\mathbb P}_1}$&$=$ & $53.007\pm0.795$  &   $c_{{\mathbb O}}$&$=$ & $0.0166\pm0.0022$          \\
$\alpha'_{{\mathbb P}_1}$&$=$& $0.5596\pm0.0078\;(GeV^{-2})$ &   $\alpha'_{{\mathbb O}}$&$=$& $0.048\pm0.0027\;(GeV^{-2})$     \\
$r^2_{{\mathbb P}_1}$&$=$& $6.3096\pm0.2522\;(GeV^{-2})$& $r^2_{{\mathbb O}}$&$=$& $0.1398\pm0.0570\;(GeV^{-2})$ \\
\hline
\multicolumn{3}{c}{\bf Pomeron$_{\bf 2}$} & \multicolumn{3}{c}{\bf $\bf \omega$-Reggeon}  \\
\hline
$\Delta_{{\mathbb P}_2}$&$=$ & $0.1669\pm0.0012$    &  $\Delta_{\omega}$&$=$ & $-0.53$ (FIXED)   \\
$c_{{\mathbb P}_2}$&$=$ & $  9.6762\pm0.1600$          &  $c_{\omega}$&$=$ & $-174.18\pm2.72$           \\
$\alpha'_{{\mathbb P}_2}$&$=$& $0.2733\pm0.0056\;(GeV^{-2})$       & $\alpha'_{\omega}$&$=$& $0.93\;(GeV^{-2})$ (FIXED)     \\
$r^2_{{\mathbb P}_2}$&$=$& $3.1097\pm0.1817\;(GeV^{-2})$   &$r^2_{\omega}$&$=$ & $7.467\pm1.083\;(GeV^{-2})$  \\
\hline
\hline
 \multicolumn{3}{c}{\bf Pomeron$_{\bf 3}$}   &  \multicolumn{3}{c}{\bf $\bf f$-Reggeon}  \\
\hline
  $\Delta_{{\mathbb P}_3}$&$=$ & $0.2032\pm0.0041$  & $\Delta_{f}$&$=$ & $-0.31$ (FIXED) \\
  $c_{{\mathbb P}_3}$&$=$ & $1.6654\pm0.0669$       &$c_{f}$&$=$ & $191.69\pm2.12$            \\
  $\alpha'_{{\mathbb P}_3}$&$=$& $0.0937\pm0.0029\;(GeV^{-2})$    & $\alpha'_{f}$&$=$& $0.84\;(GeV^{-2})$ (FIXED)    \\
  $r^2_{{\mathbb P}_3}$&$=$& $2.4771\pm0.0964\;(GeV^{-2})$&$r^2_{f}$&$=$ & $31.593\pm1.099\;(GeV^{-2})$  \\
\hline

\end{tabular}
\end{center}
\caption{Parameters obtained in Ref~\cite{threepomerons}. 
\label{tab:1}}
\end{table}}

%%%%%%%%%%%%%%%%%%%%%%%%%%%%%%%%%%%%%%%%%%%%%%%%%%%%%%%%%%%%%%%%%%%%

\end{document}